# Image registration for multi-band images taken by ONC-T onboard Hayabusa2

Toru Kouyama, Eri Tatsumi, Chikatoshi Honda, Rie Honda, Tomokatsu Morota, Yasuhiro Yokota, Shingo Kameda, Manabu Yamada, Hidehiko Suzuki, Naoya Sakatani, Masahiko Hayakawa, Yuichiro Cho, Moe Matsuoka, Kazuo Yoshioka, Hirotaka Sawada, and Seiji Sugita

**Abstract:** Hayabusa2, a Japanese sample-return mission to a C-type asteroid, arrived at its target 162173 Ryugu in June 2018. The optical navigation cameras (ONC-T, ONC-W1, ONC-W2) successfully obtained numerous images of Ryugu. ONC-T is a telescopic framing camera with a charge-coupled device (CCD), has seven filter bands in ultraviolet, visible and near infrared wavelength ranges, and were used to map the spectral distribution of the Ryugu surface. Since the locations of a target seen in ONC-T images are slightly different among different wavelength images in one multi-band observation sequence due to changes in spacecraft positions and attitudes during the filter-changing sequence, one of the image processing issues is image co-registration among images for different wavelength bands. To quickly complete the image co-registration to meet a limited mission schedule, we combined conventional image co-registration techniques with several improvements based on previous planetary missions. The results of our analysis using actual ONC-T images indicate that image co-registration can reach accuracy on the order of 0.1 pixels, which is sufficient for many spectral mapping applications for Ryugu analyses.

## 1. Introduction

Hayabusa2, a Japanese sample-return mission to near-Earth asteroid Ryugu (Watanabe et al., 2017), was launched on 3 December 2014. Hayabusa2 arrived at its target asteroid, Ryugu, on 27 June 2018, completed its entire mission program, including two touchdown operations for sample acquisitions from the Ryugu surface, left Ryugu on 13 November 2019 to return to the Earth, and successfully delivered Ryugu regolith samples (Morota et al., 2020; Tachibana et al. 2021).

Optical navigation cameras (ONCs) are the scientific and navigation instruments of Hayabusa2 and are composed of three cameras: a telescopic and multi-band camera

(ONC-T) and two wide-angle cameras (ONC-W1 and ONC-W2) (Kameda et al., 2017, Sugita et al. 2019). Because of the scientific importance of the spectral features of the Ryugu surface, which is useful for understanding the nature of Ryugu (e.g., Sugita et al., 2019; Watanabe et al., 2019; Kitazato et al., 2019; Morota et al., 2020; Okada et al., 2020; Tatsumi et al., 2021), accurate calibration for multi-band observation by ONC-T has been a key issue throughout the mission. The basic performance of ONC-T, W1 and W2 was thoroughly investigated on the ground (Kameda et al., 2015; 2017). Additionally, because any sensor can experience variation in its performance due to strong vibration during launch and after launch due to the harsh environment in space, the performance of the cameras has been continuously monitored since Hayabusa2's launch (Suzuki et al., 2018; Tatsumi et al., 2019; Kouyama et al., 2021).

Similarly to the need for accurate calibration for ONCs, accurate image registration is a key procedure to obtain a reliable spectral feature from multi-band images. Since ONC-T utilizes a filter wheel to obtain multi-band images, the positions of Ryugu surface features in a certain band image of ONC-T are unavoidably different from other bands because of the rotation of Ryugu and variations in spacecraft position and attitude. If there is even one- or few-pixel misregistration, we may seriously misrecognize spectral features in a region containing complex surface textures.

Because of the importance of the procedure, image co-registration software for planetary and small asteroid exploration has been developed and widely used, such as Integrated Software for Imagers and Spectrometers (ISIS) by the United States Geological Survey (USGS) [cf. Sides et al. (2017); Backer et al. (2018)]. In the Hayabusa2 mission, ONC-T conducted numerous observations with various distances, such as observations from a 20-km distance (so-called home position, HP) for global mapping and observations from less than 100 m of distance to capture the details of the Ryugu surfaces with high resolution (better than 1 cm/pix) for safe landing. The range of the difference notably varied in different observation sequences from a few pixels at HP observations to several hundred pixels at low-altitude observations.

Due to observations at various altitudes, an efficient application for image co-registration was required, which can handle images taken at various distances to a target from less than 100 m to 20 km. In addition, the software had to be robust for use with only a single or few parameter settings during the entire mission period to minimize the time for investigating appropriate parameter settings, which was critical for making prompt decisions to select landing sites.

To address these issues in the Hayabusa2 mission, we developed our own image registration application, which can work with a few parameter settings during the mission period and is based on heritages from previous planetary and remote-sensing missions with insight based on computer vision techniques. In Section 2, we present an overview of the algorithm and calculation flow of the developed procedure of image registration. In Section 3, we present examples of the image co-registration with actual images taken by ONC onboard Hayabusa2 and examine the performance of the developed application. In Section 4, we provide conclusion remarks and discuss future applications.

## 2. Image registration applied to ONC-T images
### 2.1 Overview of image registration procedure

To investigate the surface spectrum of Ryugu from ONC-T multi-band images, accurate image registration is essential to avoid errors from misregistration. To achieve accurate image registration (i.e., subpixel-level image registration), we combined several image registration techniques.

One key technique is to repeat template matching and affine transformation three times based on the idea of a coarse-to-fine approach (Tanimoto, and Pavlidis, 1975; Rosenfeld and Vanderbrug, 1977) with subpixel estimation by interpolating a cross-correlation surface (cf. Ogohara et al., 2011). Another important technique is feature-based matching (e.g., Scale-Invariant Feature Transform (SIFT) (Lowe, 2004); Speed Up Robust Feature (SURF) (Bay et al., 2008)). This technique can roughly match images even if they have a large displacement, e.g., several hundred pixels, with a small computational cost. Feature-based matching is adopted before conducting the coarse-to-fine approach of image registration for low–elevation images. For such low-elevation images, displacement of the same feature in two different band images sometimes reaches half of the ONC-T image frame.

Figure 1 shows the calculation flow of the coarse-to-fine approach in the ONC-T image registration. During the co-registration, a master band image (or reference band image) was determined for each multi-band observation. In the ONC-T pipeline, a band taken in the middle of an image sequence is defined as a reference band. Table 1 shows the temporal imaging order and defined master band.

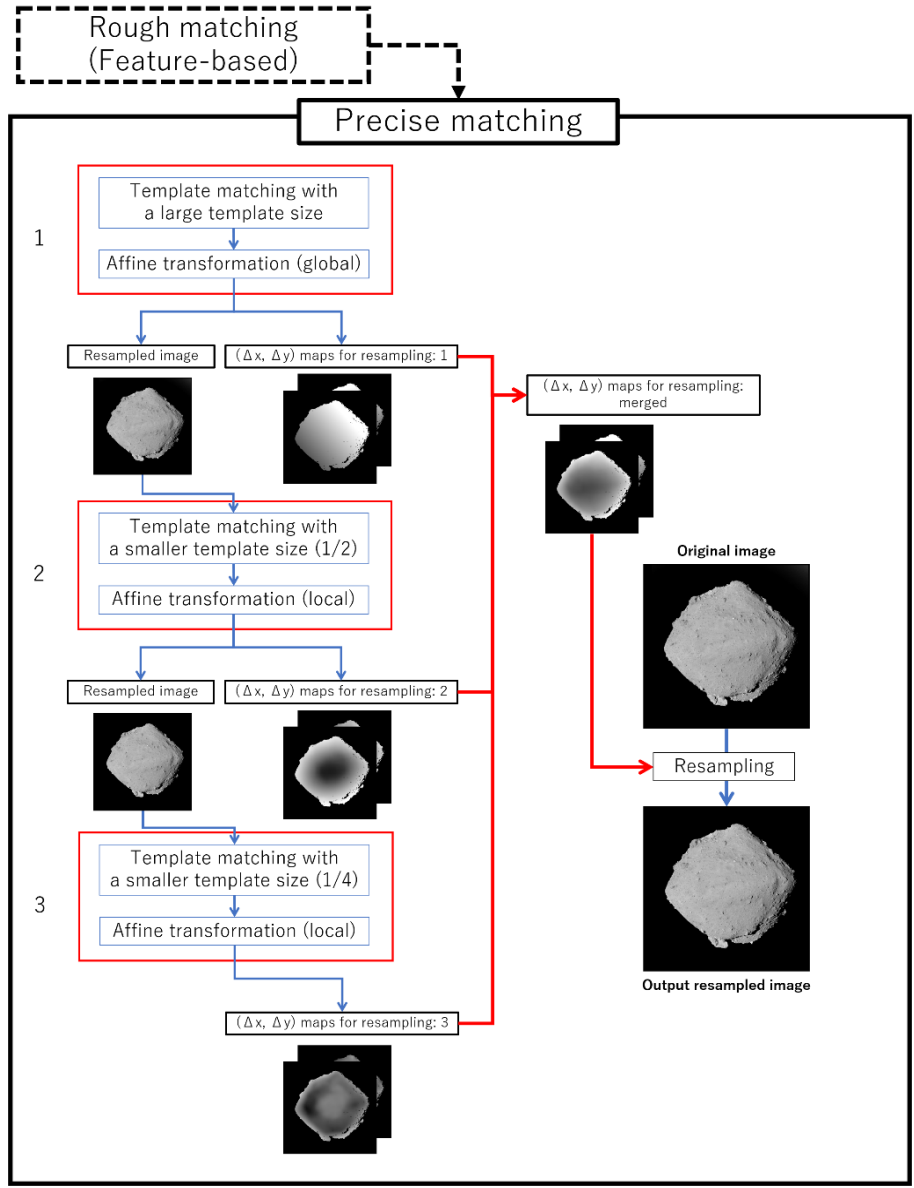

**Figure 1.** Calculation flow of image registration for ONC-T multiband images in the ONC-T pipeline.

**Table 1**. Temporal image sequence and master band for co-registration.

| Number of bands in image sequence | Temporal order of filters in image sequence | master band image |
|---|---|---|
| 3 | {v, ul, x} | ul |
| 4 | {v, ul, b, x} | ul |
| | {v, w, x, ul} [1] | x |
| 6 | {v, w, x, p, b, ul} | x |
| 7 (nominal) | {v, w, x, Na, p, b, ul} | Na |

[1] This band set is used only for the date 20190207.

**2.2 Rough matching based on feature-based matching**

Figure 2 shows an example of a feature-based matching adopted before conducting the coarse-to-fine approach for low–elevation images. First, we found the positions of matched pairs between two images (Figure 2a). In this time, the scale-invariant similarity estimated by a feature matching algorithm is used to find possible feature pairs. Then, we performed affine transformation with a single affine parameter to obtain roughly matched image pairs (Figure 2b). In our method, SURF was used as a feature matching algorithm, and RANdom SAmple Consensus (RANSAC) (Fisher and Bolles, 1981) was applied to eliminate obvious outliers in the matched pairs obtained from SURF, which is a key procedure to enhance the robustness of our method.

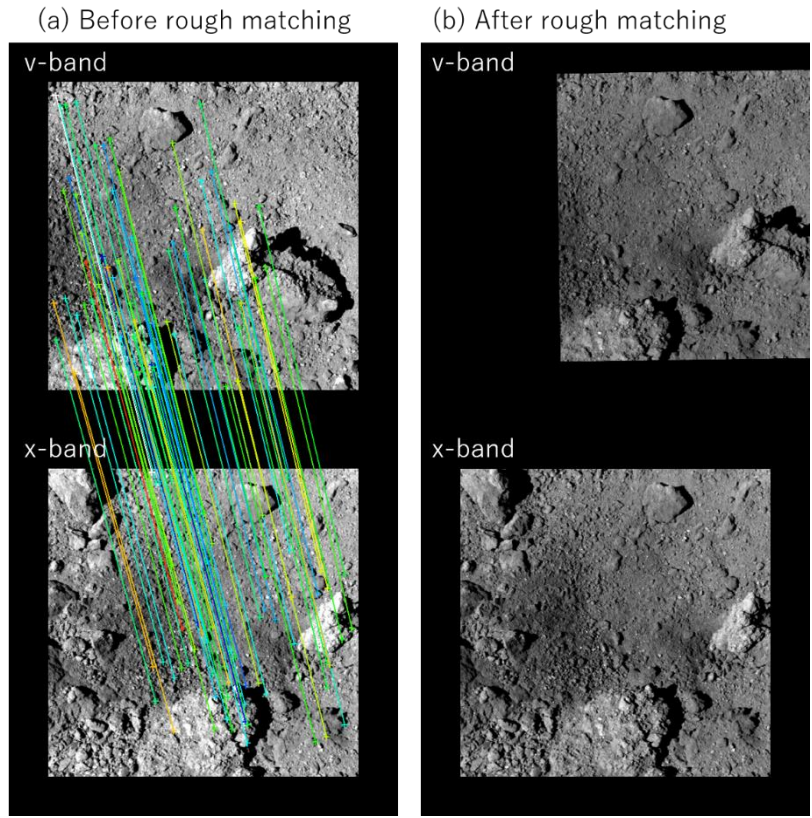

**Figure 2.** (a) Example of feature-based matching for v-band and x-band images of low-elevation observations (June 13, 2019). The lines indicate pairs of matched features. The color represents the similarity of a matched pair (blue to red). (b) After performing affine transformation as rough matching based on the positions of the matched pairs in (a).

**2.3 Precise matching based on template matching**

Template matching is conducted to obtain the so-called "optical flow" vectors in two different band images, as illustrated in Figure 3. At the first image registration step, which is for coarse matching, a template size of 129 x 129 pixels is used, and templates are set with an interval of 64 pixels in both *x* and *y* directions. For the second and third steps, half- and 1/4-size templates are used with shorter (half and 1/4) intervals, respectively. Template matching is performed based on cross-correlation between a template and a search area, where a peak position of the cross-correlation surface is considered to represent the displacement of two images. Then, the correlation surface around the peak position is interpolated with a hyperboloid function to estimate a subpixel position of the matched point (cf. Ogohara et al., 2012). Finally, affine transformation is conducted to adjust an image of a target band to an image of a reference band based on subpixel-level optical flow vectors.

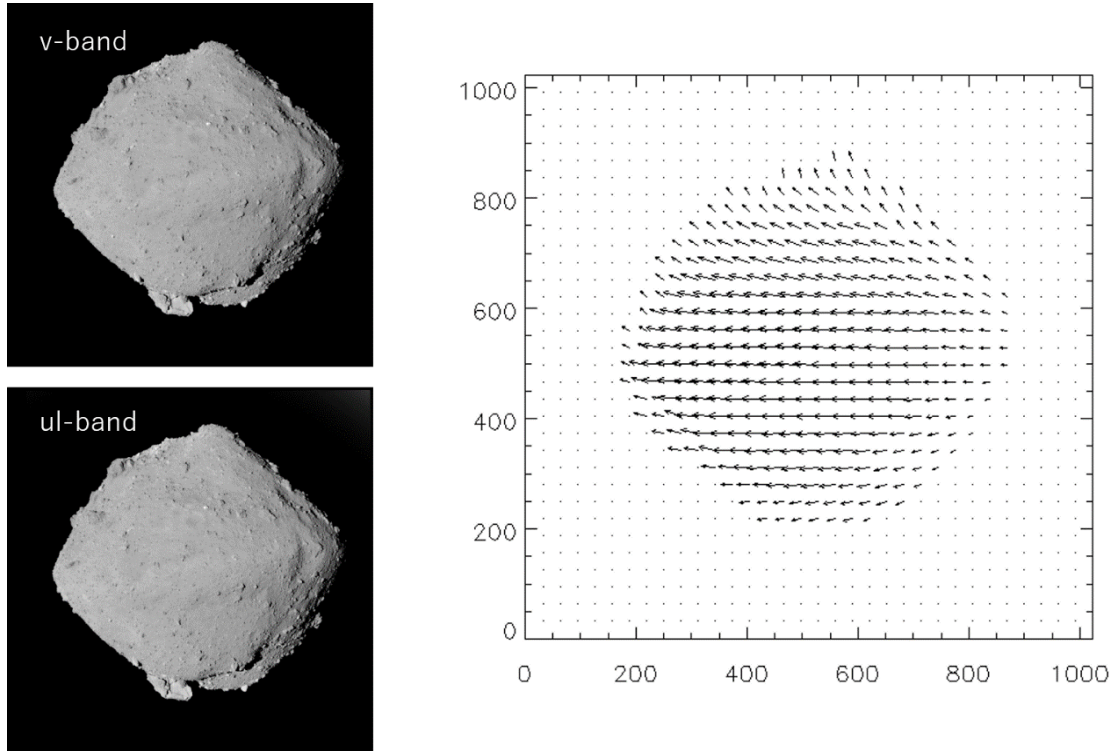

**Figure 3.** An example of optical flow vectors obtained from template matching between v-band (a reference band) and ul-band (a target band) images.

Since the positions of a Ryugu surface feature in two images are expected to be largely different in the first step, one affine matrix derived from entire optical flow vectors is used for the affine transformation of the entire image frame. In the second and third steps, local affine transformation is conducted at every position of an optical flow vector by using a local affine matrix because nonuniform transformation is required to match two band images with a subpixel level due to the complicated Ryugu surface shape. In the local affine transformation, only 9 vectors (a vector at a target template position and 8 surrounding vectors) are used to estimate a local affine matrix, and an affine transformation is conducted within only a region in the target template. The measured displacements in the x and y directions (Figure 4) to match images are also stored at every affine transformation step.

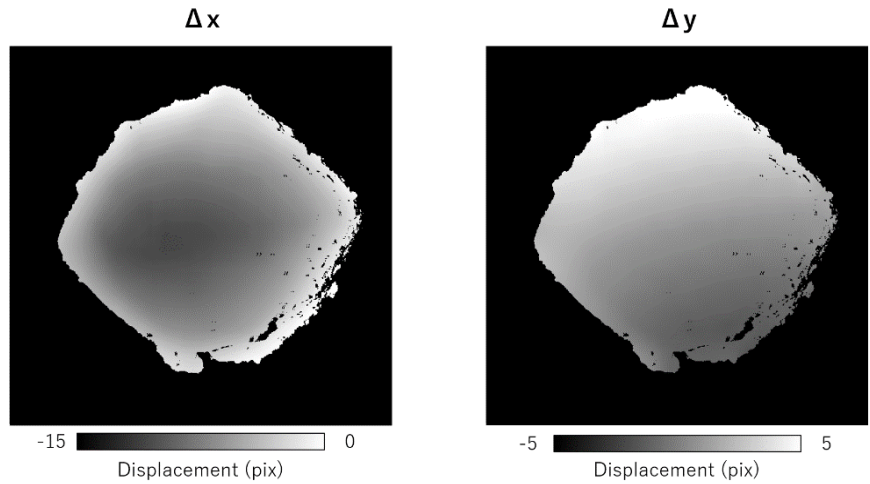

**Figure 4.** Maps of the measured displacement (Δx, Δy) between target and reference band images in Figure 3.

There can be optical flow vectors that indicate wrong directions due to the lack of features in a template area and/or effect from image noise, which reduce the correlation coefficient for matching. Since such incorrect vectors provide an unnatural result of resampling, they should be replaced with more reasonable vectors.

Because there is basically no moving object on the Ryugu surface (except for effects from touchdown operations and impactor operation of Hayabusa2), the direction of an optical flow vector should be very similar to those from nearby surrounding vectors. Considering this characteristic, we implemented a procedure that finds an incorrect vector by comparing a vector direction with the directions of surrounding vectors and by checking the magnitude of correlation of template matching. If the direction of a focused vector is sufficiently different from those of surrounding vectors or has a smaller correlation than a threshold, then we replace the direction of the vector with the mean direction of surrounding vectors. We set thresholds of 0.5 for a correlation value to decide whether the vector is reliable and 2 pixels for a vector direction to determine whether its vector direction is similar to those of surrounding vectors.

Finally, the "original" target image is adjusted to the reference image with bilinear interpolation using the merged Δx and Δy maps (see Figure 1) after three affine transformations. This procedure aims to perform resampling only once to avoid as much image blurring due to resampling as possible.

## 3. Evaluation of the image registration performance
### 3.1 Examples of image registration

Figure 5 shows a comparison of color-composite images (R: w-band, G: v-band, B: b-band) before and after image registration. Because of the motion of Ryugu's rotation during multiband observations, artificial color features appeared on boulders in the image before image registration, whereas there were fewer color features (i.e., similar surface reflectance among wavelengths) after image registration, which is consistent with previous reports (cf. Kitazato et al., 2019; Tatsumi et al., 2020).

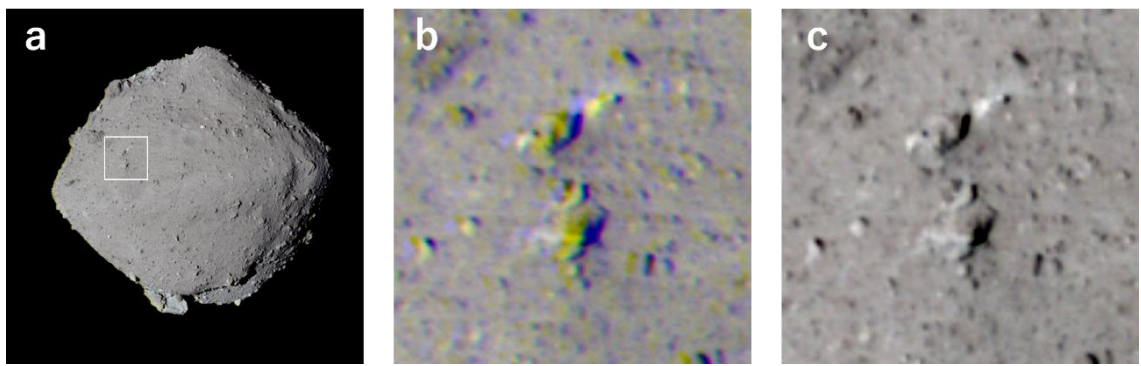

**Figure 5.** (a) Color-composited Ryugu images taken by ONC-T (R: w-band, G: v-band, B: b-band) after applying the image registration. (b) (c) Close-up images of the Ryugu surface indicated by a white rectangle in (a), (b) before and (c) after applying the image registration.

As another example, image co-registration was adopted for multi-band Earth images that were obtained after an Earth swing-by operation of Hayabusa2 on December 5, 2015. Because the position and attitude of Hayabusa2 gradually varied during an imaging sequence, the Earth positions in ONC-T FOV were slightly different in different band images.

Due to this slight difference, a band ratio image (e.g., a pair of w-bands and x-bands that is suited for evaluating the vegetation existence) appeared to be a type of shape-enhanced image (Figure 6a). After the image co-registration, such artificial enhancement vanished, and the shapes of continents were easily confirmed (Figure 6b).

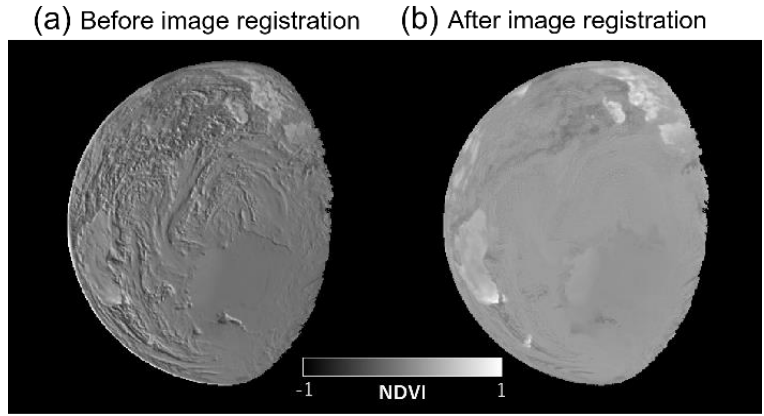

**Figure 6.** Normalized Difference Vegetation Index (NDVI) measured from Earth images taken by ONC-T (R: w-band, IR: x-band, NDVI = (IR-R)/(IR+R)) (a) before and (b) after the image registration procedure.

Figure 7 shows an example of image registration results for images of low-elevation observations, for which we adopted a rough matching procedure before precise image registration. In the example, although only ~1/3 of the region in the ONC-T FOV overlapped among the three band images, our method successfully and appropriately matched images with no special parameter settings and with a small computational cost for SURF.

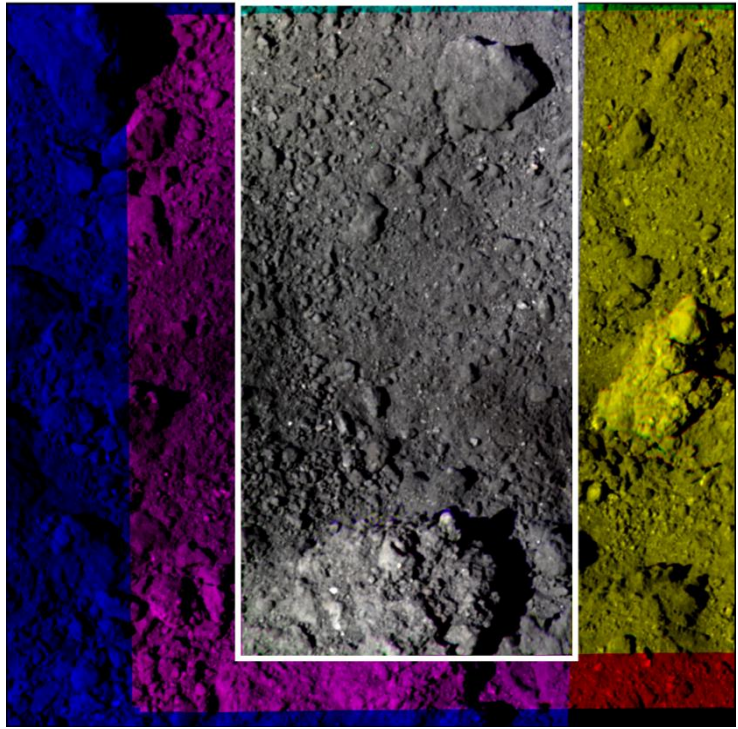

**Figure 7.** Color-composited Ryugu image taken by ONC-T (R: w-band, G: v-band, B: b-band) after applying the image registration for a low-elevation multiband observation taken on June 13, 2019. Because of the large FOV difference among the bands due to spacecraft motion during the multi-band observation, only a region indicated by a white rectangle overlapped among the three bands.

### 3.2 Accuracy of image registration

To evaluate the accuracy of the image registration, we artificially shifted an Earth image with subpixel values in both $x$ and $y$ directions and applied image registration with the original image and shifted image. In the $x$ direction, we simply shifted the image by 0.5 pixels. In the $y$ direction, we shifted the image with $\sin(y/256 \times 2\pi)$ pixels to evaluate possible local motions. Figure 8 shows the result of the magnitude of displacement measured by the image registration. We confirmed that the measured mean displacement in the $x$ direction was $0.5 \pm 0.08$ pixels. Similar performance (0.1-pixel error) was also confirmed in the y direction, which indicates that our method can match images with an accuracy on the order of 0.1 pixels.

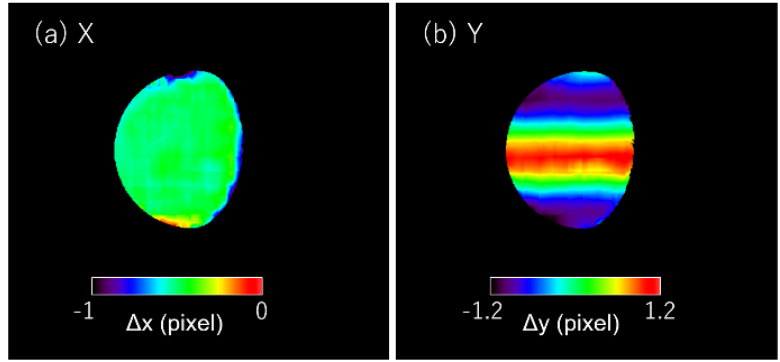

**Figure 8**. Derived displacement pixels in the (a) x and (b) y directions using an artificially shifted Earth image. The given displacements were 0.5 pixels in the x direction and sin(y/256 × 2π) in the y direction.

## 4. Summary

In this study, we developed a robust and accurate method of image registration that combines coarse-to-fine template matching with subpixel estimation and feature-based matching to suppress additional computational costs for images of low-elevation observations. Although each technique is conventional or previously developed, the combination of the techniques provides sufficient robustness, sufficient accuracy (0.1-pixel level) and reasonable computational time. Because of the robustness, our method can work in the ONC-T data pipeline during the entire rendezvous phase around Ryugu, including HP observations and low-elevation observations, with few parameter settings. Since image co-registration is a fundamental process for any multi-band imaging, the developed method can be applied in future missions. Our method in this study can be applied to image pairs with various resolutions. We believe that the method can be a fundamental tool in future sample return missions, which involves remote sensing with far distances and touch down or landing operations so that the image resolution can significantly vary, similar to the Hayabusa2 mission. More optimization of computational efficiency and robustness should be a key issue in future studies.

**Acknowledgment:**
We thank the Hayabusa2 project team for achieving the science observations. This study is supported by the following grants: JSPS KAKENHI 19K14789 and 20H01958, and JSPS Core-to-core program "International Network for Planetary Sciences".